\documentclass[preprint,superscriptaddress]{revtex4}

\usepackage{graphicx}
\usepackage{color}
\usepackage{url}
\usepackage{gensymb}
\usepackage{epsf}

\begin{document}

\title{Disentangling Cooper-pair formation above $T_{\rm c}$ from the pseudogap state in the cuprates}

\author{Takeshi~Kondo}
\affiliation{Ames Laboratory and Department of Physics and Astronomy, Iowa State
University, Ames, IA 50011, USA}
\author{Yoichiro~Hamaya}
\affiliation{Department of Crystalline Materials Science, Nagoya University, Nagoya
464-8603, Japan}
\author{Ari~D.~Palczewski}
\affiliation{Ames Laboratory and Department of Physics and Astronomy, Iowa State
University, Ames, IA 50011, USA}
\author{Tsunehiro~Takeuchi}
\affiliation{Department of Crystalline Materials Science, Nagoya University, Nagoya
464-8603, Japan}
\affiliation{EcoTopia Science Institute, Nagoya University, Nagoya 464-8603, Japan}
\author{J.~S.~Wen}
\affiliation{Condensed Matter Physics and Materials Science Department, Brookhaven
National Laboratory, Upton, New York 11973, USA }
\author{Z.~J.~Xu}
\affiliation{Condensed Matter Physics and Materials Science Department, Brookhaven
National Laboratory, Upton, New York 11973, USA }
\author{Genda~Gu}
\affiliation{Condensed Matter Physics and Materials Science Department, Brookhaven
National Laboratory, Upton, New York 11973, USA}
\author{J\"org~Schmalian}
\affiliation{Ames Laboratory and Department of Physics and Astronomy, Iowa State
University, Ames, IA 50011, USA}
\author{Adam~Kaminski}
\affiliation{Ames Laboratory and Department of Physics and Astronomy, Iowa State
University, Ames, IA 50011, USA}

\maketitle

 {\bf  The discovery of the pseudogap in the cuprates created significant excitement amongst physicists as it was believed to be a signature of pairing \cite{Emery}, in some cases well above the room temperature. In this "pre-formed pairs" scenario, the formation of pairs without quantum phase rigidity occurs below $T^*$. These pairs condense and develop phase coherence only below $T_{\rm c}$ \cite{Emery}.
In contrast, several recent experiments reported that the pseudogap and superconducting states are characterized by two different energy scales \cite{ShenScience,Twogap,EricTwogap,Raman}, pointing to a scenario, where the two compete\cite{HongCompetition,KondoCompetition,Rustem}. 
However a number of transport, magnetic,  thermodynamic and tunneling spectroscopy experiments consistently detect a
signature of phase-fluctuating superconductivity above $T_{\rm c}$\cite{OngPRB,OngDiamagnetism,Terahertz,CriticalField,Josephson,DavisFingerprint} leaving open the question of whether the pseudogap is caused by pair formation or not.
Here we report the discovery of a spectroscopic signature of pair formation and demonstrate that in a region of the phase diagram commonly referred to as the ``pseudogap", two distinct states coexist: one that persists to an intermediate temperature $T_{\rm pair}$ and a second that extends up to $T^*$. 
The first state is characterized by a doping independent scaling behavior and is due to pairing above $T_{\rm c}$, but significantly below T*. 
The second state is the ``proper" pseudogap -  characterized by a ``checker board" pattern in STM images, the absence of pair formation, and is likely linked to Mott physics of pristine CuO$_2$ planes. 
 $T_{\rm pair} $  has a universal value around 130-150K even for materials with very different $T_{\rm c}$, likely setting limit on highest, attainable Tc in cuprates. The observed universal scaling behavior with respect to $T_{\rm pair} $  indicates a breakdown of the classical picture of phase fluctuations in the cuprates.}

The traditional approach of exploring the paring origin above $T_{\rm c}$
by tracking the energy scale of spectral features has not yielded convincing
results so far, as these features are poorly defined above $T_{\rm c}$ due to
broad spectral peaks. 
The apparent smooth evolution of the spectral gap from the lowest temperatures up to $T^*$\cite{HongPseudogap,LoeserPseudogap,ZhouBi2201,Nakayama}
has previously been interpreted as key evidence for a common origin of the pseudgap and
pairing gap. However, very detailed, high precision data (e. g. Fig. S3E in 
Supplementary Information), shows the gap size varies non-monotonically across $T_{\rm c}$. 
This behavior raises doubt about the above interpretation. 
A better approach is to investigate the spectral weights, which are easier to quantify and
in many cases interpret. A key such measure is the density of states at the
Fermi energy D($E_{\rm F}$). In conventional, clean superconductors this
weight is zero below $T_{\rm c}$, but can be finite if there are strong impurity
scattering effects. In such cases D($E_{\rm F}$) reflects the pair breaking state.
Another possibility is the case of an inhomogeneous superconductor such as cuprates \cite{STMInhomogeneity,Yazdani}, 
where superconducting and "normal" patches coexist in the sample, with the latter being likely due to pair breaking states
states (generic density wave states, localization etc).
In either case, D($E_{\rm F}$) is proportional to the number of "normal" electrons present at $E_{\rm F}$ due to pair breaking processes.
Measurements of the temperature dependence of this fundamental, yet rarely explored quantity enabled us to disentangle the electronic ground states of cuprates. Since the spectral gap in the cuprates displays significant momentum dependence, in our study we used the intensity of the spectral function at $E_{\rm F}$, I($E_{\rm F}$, $k$), which integrated over all momenta, equals D($E_{\rm F}$) (modulo matrix elements). This approach allowed us to isolate the  behavior at a specific k-point and avoid smearing due to averaging (since in general the temperature dependence of  I($E_{\rm F}$, $k$) will vary with momentum.)

In Fig.1a-c we examine the temperature evolution of the spectral
line shape measured at the antinodal Fermi momentum in optimally doped
Bi2212 ($T_{\rm c}$=90K). Symmetrized EDCs\cite{NormanNature} 
show the opening of  pseudogap on cooling below $T^*$ ($\sim
$210K).  As the temperature is decreased below $T_{\rm c}$, a superconducting coherent peak
appears.  We obtain the spectral changes with temperature by subtracting the spectrum at the highest
temperature from all the spectra measured, as shown in Fig. 1b. 
Now we will focus on the loss of spectral weight close to the Fermi
level, $W(E_{\rm F})$) (hatched area in Fig. 1b), which is due to the previously-discussed "normal" electrons.
The temperature dependence  of  $W(E_{\rm F})$ is plotted in Fig 1d. 
On cooling through $T^*$, the spectral weight decreases linearly, which is a
characteristic behavior of the pseudogap state. An astonishing feature seen in this plot is a clear deviation from the linear behavior  (indicated by an arrow).  Since the temperature dependence below and above this point is very different the arrow marks the onset of another distinct state.
The onset temperature $T_{\rm pair} (\sim$150K) of this transition is considerably higher than $T_{\rm c}$ (=90K), but is also significantly lower that the pseudogap temperature T* ($\sim$210K).
This state likely arises due to the pairing of the electrons  because the weight loss associated with this state smoothly evolves through $T_{\rm c}$. If we extrapolate the linear variation of $W(E_{\rm F})$ down to T=0, we obtain approximate values of the spectral weight lost due to pseudogap, $W_{\rm PG}$ (blue area) and pairing $W_{\rm pair}$ (red area) as marked in Fig. 1d.

We now verify this hypothesis by studying how these quantities vary with doping in related samples of Bi2201, where $T_{\rm c}$ and $T^*$ are more separated over a wide range of carrier concentrations \cite{NMR}. 
The  first row of Fig. 2a-g show symmetrized EDCs measured at the antinode for various temperatures and dopings
from underdoped (left side) to overdoped (right side) samples. 
The spectral changes with temperature (obtained by subtracting the high temperature spectrum from data in panels a-g) are plotted in the
panels h-n.  We will focus again on the spectral weight close to the Fermi level, $W(E_{\rm F)}$) (see Fig1b), which is plotted in panels o-u for all samples.  As in case of Bi2212 (Fig. 1) $W(E_{\rm F)}$ is linear below T* at high temperatures, then suddenly deviates from a straight line - defining a new temperature scale $T_{\rm pair}$. The temperature dependence of $W(E_{\rm F)}$) evolves in a surprisingly systematic manner with doping. The linear part becomes longer with underdoping, as both T* and $T_{\rm pair}$ increase. Eventually, at the lowest dopings, $W(E_{\rm F})$ is linear down to the lowest temperature, because the pseudogap completely dominates the spectra and prevents the formation of the superconducting peak at the antinode in highly underdoped samples \cite{KondoCompetition}. $W_{\rm pair}$ decreases and $W_{\rm PG}$ increases with underdoping, a behavior that is consistent with the competing nature of the pairing and the pseudogap. 

To validate our assertion about the pairing origin of $W_{\rm pair}$, we extract this quantity for each doping by subtracting the interpolated  $W_{PG}(T)$ line from each of the $W(E_{\rm F},T)$ curves and compare  them in Fig. 3. Obviously the magnitude and onset temperature of $W_{\rm pair}$ is very different for each doping. To make a fair comparison,we rescale the vertical axis for each curve by its maximum value at T=11K and horizontal axis by $T_{\rm pair}$. Surprisingly the curves for all dopings fall on top of each other, demonstrating a universal scaling of $W_{\rm pair}$, that smoothly evolves through $T_{\rm c}$.
This almost perfect scaling behavior not only validates our extraction of the  paring component, but it also gives compelling evidence that the pseudogap origins are not due to paring. 

We summarize our results in Fig. 4 in the form of a phase diagram, where we plot the spectral weights $W_{\rm pair}$, $W_{\rm PG}$ using a color scale and and onset temperatures of pair formation $T_{\rm pair} $ and the peudogap $T^*$, extracted using our ARPES analysis for all dopings. We compare our data to the detection of phase-fluctuating superconductivity by magnetization\cite{OngDiamagnetism}, the Nerst effect\cite{OngPRB} and NMR\cite{NMR} as well as the pseudogap temperature extracted from resistivity measurements performed on the same samples. (see Fig. S1 of Supporting Online Material).
We note the excellent agreement between these probes and ARPES. Note that ARPES
reports slightly higher onset temperatures, due to fact that the other probes
are sensitive to an average over larger portions of the Fermi momenta, while
with ARPES we can extract these directly for the antinodal areas only, where these
temperatures are largest. The paring temperature ($T_{\rm pair} $) (panel b)  increases steadily from the overdoped side of the phase diagram towards optimal doping. For dopings lower than optimal,  it levels off at $\sim$130K. 
This behavior contrasts to that of the pseudogap temperature ($T^*$) (panel c), which monotonically increases up to the lowest doping.  It is quite surprising that the paring temperature of Bi2201 is almost the same as that of Bi2212 [$\sim$150K, see Fig. 1d], despite large difference of $T_{\rm c}$ for these two systems. This strongly suggests that, although the $T_{\rm c}$ has a significant variation for different types of cuprates,  the onset temperature of paring (130K-150K) is universal and similar to highest achievable superconducting temperature in the cuprates. 

Quantitive analysis of the very detailed ARPES data presented here provides clear evidence for a spectroscopic temperature scale $T_{\rm pair} $ ,distinct from $T_{c}$ and $T^{\ast }$. It demonstrates that pairing and the pseudogap are two fundamentally different, coexisting and competing states. The competition between the two states plays a key role in the determination of $T_{\rm c}$, where bulk superconductivity is established. 
The doping dependence of the pseudogap weight ($W_{\rm PG}$), resembles that of the ``checker board" pattern\cite{Hanaguri}, both of them become more pronounced in the underdoped region\cite{EricCDW} of the phase diagram. 
It is very likely that both the pseudogap and  ``checker board" pattern have a common origin and are due to an ordered state.
This is strongly supported by recent STM/STS results for the same Bi2201 samples\cite{EricFS} as in our study, showing 
that the energy scale of the ``checker board" pattern is almost identical over a wide range of doping with that of the pseudogap observed in our ARPES data.
Our ARPES results have important consequences for understanding the mesoscale properties in the cuprates. 
Below $T^*$ an ordered state emerges, that is  likely a result of the underlying Mott physics and manifests itself as a "checker board" pattern \cite{Hanaguri}. 
Upon cooling below $T_{\rm pair}$, the pairing and local superconductivity emerges at locations, where the order parameter of the pseudogap is suppressed due to impurities or defects. 
These superconducting inhomogeneities are observed \cite{STMInhomogeneity}, and give rise to diamagnetic\cite{OngDiamagnetism} and Nernst \cite{OngPRB} signals. On further cooling, the superconducting order parameter increases and an inhomogeneous superconducting state emerges in over larger and larger parts of the sample, which gives rise to an accelerated loss of spectral weight at $E_{\rm F}$ as observed in our ARPES data.  
Bulk superconductivity and coherent quasiparticle peaks emerge below $T_{\rm c}$, where the pair scattering in the system is suppressed. 
The fact that we can observe perfect scaling of this quantity with temperature, regardless of the impurity content and doping, implies that the only two relevant energy scales are the order parameter of the pseudogap state and the magnitude of the pairing potential, making the existence of the three temperature scales $T_{\rm c}<T_{\rm pair} <T^{\ast }$, a universal aspect of underdoped cuprates. 
This scaling behavior of $W_{\rm pair}$ gives an important clue as to the nature of the Cooper-pair fluctuations: it is not sufficient to solely consider phase fluctuations of the pairing field. $T_{\rm pair} $ is the temperature where the amplitude of the pairing field melts, i.e. where the strength of incoherent pairing disappears.  The emergence of $T_{\rm pair}$ as the relevant temperature of the scaling plot requires that the pair-amplitude and phase fluctuation are equally crucial below $T_{\rm pair}$. 
A pairing interaction that is in the extreme strong coupling limit, reminiscent of Mott physics,  was shown to lead to simultaneous amplitude and pair modes that separate coherent and local pairing\cite{ChubukovAmplitudeandPhase}.

\textbf{Methods} 

Optimally doped Bi$_2$Sr$_2$CaCu$_2$O$_{8+\delta}$ (Bi2212) single crystals with $T_{\rm c}$=90K (OP90K) and 
(Bi,Pb)$_2$(Sr,La)$_2$CuO$_{6+\delta}$ (Bi2201) single crystals with various $T_{\rm c}$'s were grown by the conventional 
floating-zone (FZ) technique.  (see Supplementary Information on the sample characterization) To  precisely analyze the ARPES  spectra, we partially substituted Pb for Bi for all doping samples to suppress the modulation in the BiO plane,  which usually contaminates the ARPES signal.
ARPES data was acquired using a laboratory-based system consisting of a Scienta SES2002 electron analyzer and GammaData Helium UV lamp. All data were acquired using the HeI line with a photon energy of 21.2 eV. The angular resolution was $0.13^\circ$ and $\sim 0.5^\circ$ along and perpendicular to the direction of the analyzer slits, respectively.  The energy resolution was set at $\sim10$meV - confirmed by measuring the energy width between the 90$\%$ and 10$\%$ of the Fermi edge from the same Au reference. 
Custom designed refocusing optics enabled us to accumulate  high statistics spectra in a short time without being affected by possible sample surface aging.
In the analysis we used symmetrized EDCs normalized over the whole energy range (-0.4eV $\leq E\leq $ 0.4eV) for each spectrum.  
We have verified that a particular normalization scheme does not affect the results of our analysis. (see Supplementary Information on the details)

\textbf{Acknowledgements}

We thank Andy Millis, Chandra Varma and Mike Norman for useful discussions. 
This work was supported by Basic Energy Sciences, US DOE. The Ames Laboratory is
operated for the US DOE by Iowa State University under Contract No. W-7405-ENG-82.

{\bf Author Contributions}

 T.K. and A.K. designed the experiment. T.K., Y.H., T.T., J.S.W, G.Z.J.X, and G.G grew the high-quality single crystals. T.K. and A.D.P acquired the experimental data and T.K. performed the data analysis. T.K., A.K. and J.S. wrote the manuscript. All authors discussed the results and commented on the manuscript. 

\bigskip\

\bigskip

\begin{figure*}[tbp]
\includegraphics[width=6in]{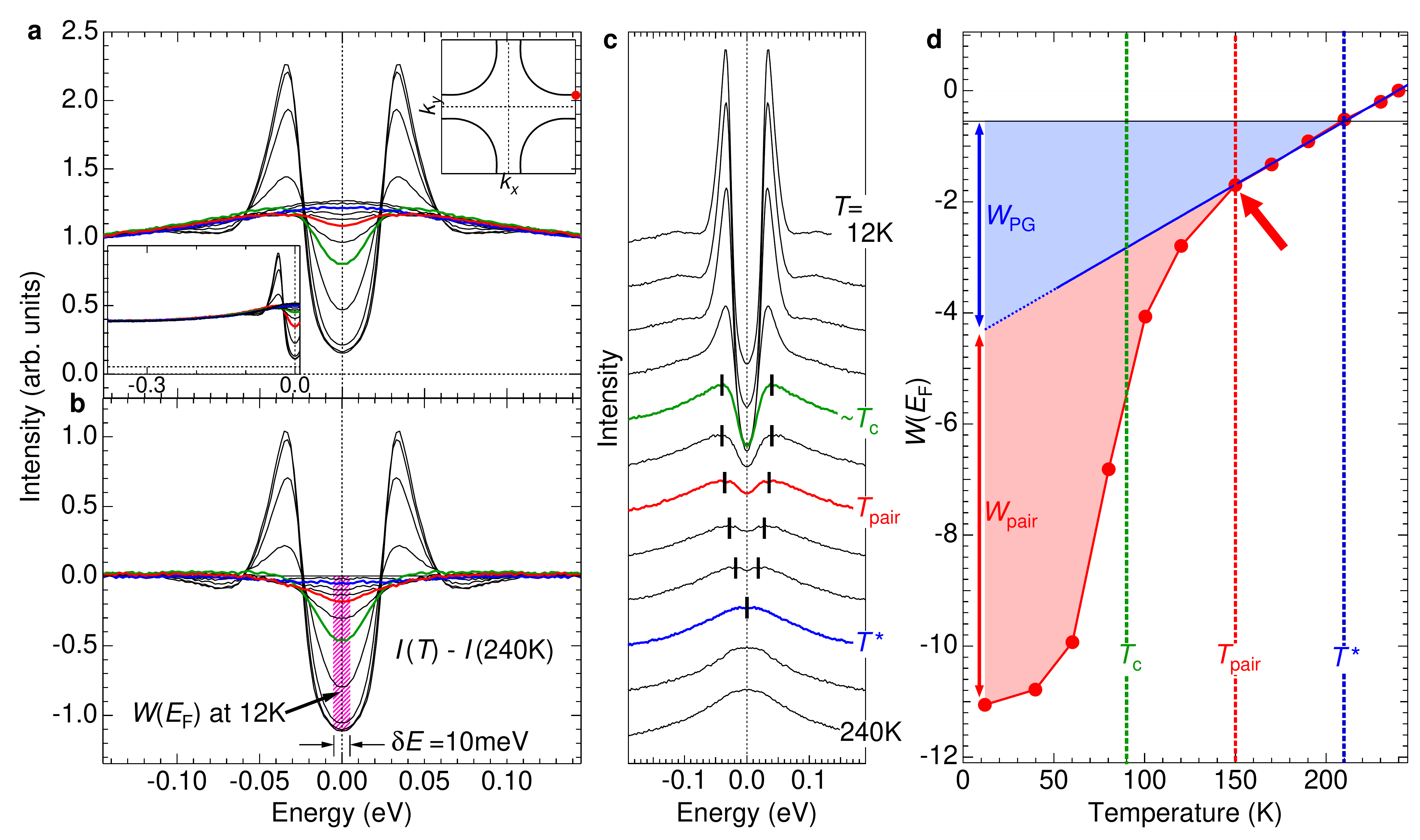}
\caption{ Temperature dependence of the spectral weight at the chemical potential.  {\bf a} Symmetrized energy distribution curves (EDCs) \cite{NormanNature}
for various temperatures from deep
below $T_{\rm c}$ to above the preusogap temperature ($T^*$). The spectra were measured at the
antinode in optimally doped Bi2212 ($T_{\rm c}$=90K). 
{\bf b} Difference spectra: the spectrum measured at the highest temperature is subtracted from each of spectra in (a). 
Spectral weight close to the Fermi level ($W(E_{\rm F}$), hatched area) is
estimated by integrating the spectral intensity in (b) within an energy
window of the experimental energy resolution (10meV). {\bf c} Same spectra as in A
with offsets. Spectral gaps are indicated with bars. {\bf d} The temperature
dependence of $W(E_{\rm F}$). The paring temperature,
$T_{\rm pair} $, is defined as the onset temperature of deviation  (marked by arrow) from 
a linear behavior seen at higher temperatures. 
The pseudogap temperature, $T^*$, is defined to be the temperature where the two spectral peaks in
the symmetrized EDCs merge into a single peak as seen in (c). The three temperatures, $T_{\rm c}$ (green),
$T_{\rm pair}$ (red), and $T^*$ (blue) are indicated with dashed lines. The paring
weight (Wpair, red area) and the pseudogap weight ($W_{\rm PG}$, blue area) are
separated by a line extrapolated from the linear behavior of $W(E_{\rm F}$) at high
temperatures.}
\label{fig1}
\end{figure*}

\begin{figure*}[tbp]
\includegraphics[width=7in]{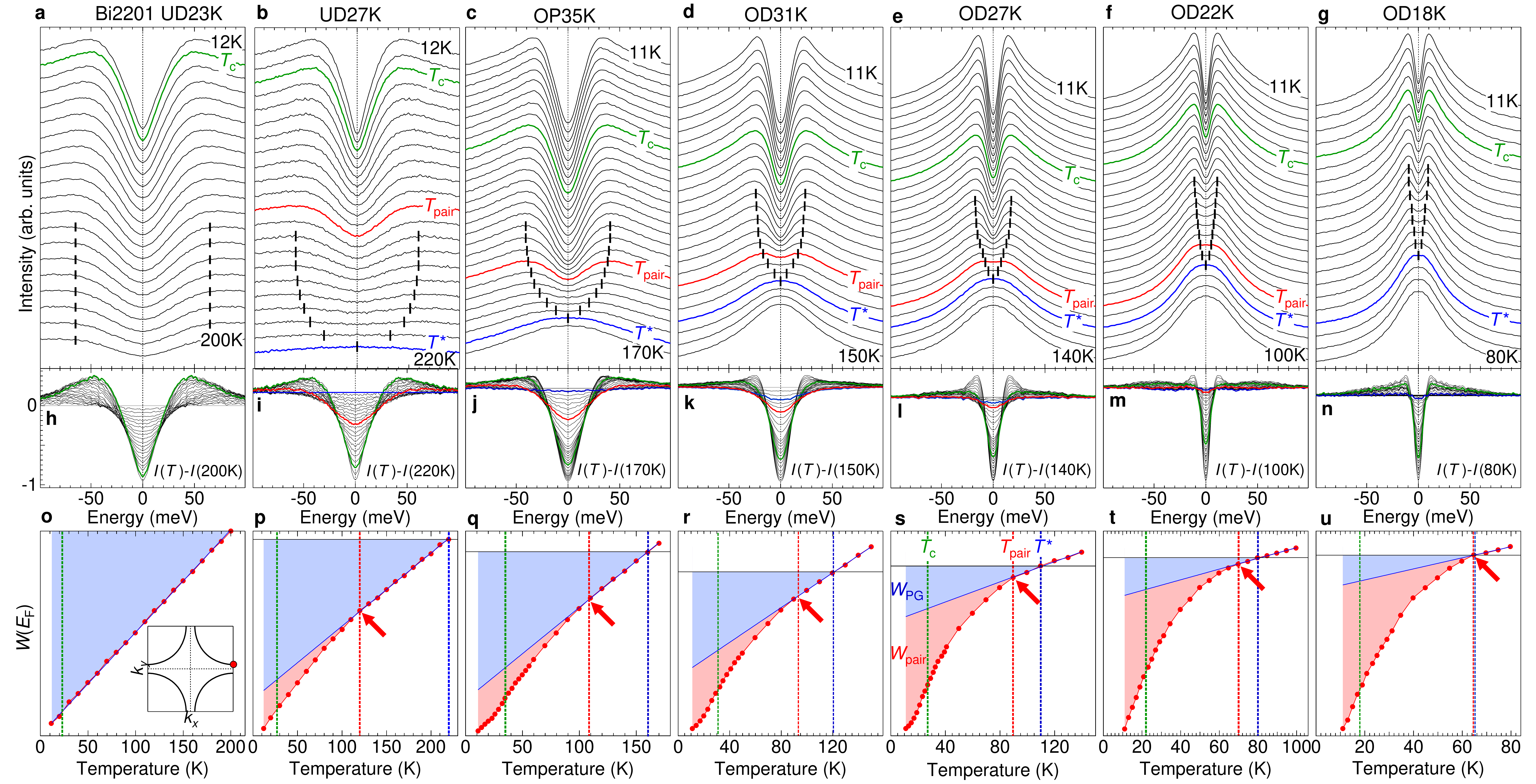}
\caption{Doping and temperature dependence of spectral weight at the chemical potential {\bf a-g} Symmetrized EDCs at various temperatures measured at the antinode in Bi2201 over a wide range of doping from the underdoping (left) to overdoping (right). An offset is used for clarity.
{\bf h-n} Difference spectra: the spectrum at the highest temperature is subtracted from each of the spectra in a-g. 
{\bf o-u} Temperature dependence of the spectral weight close to $E_{\rm F}$, $W(E_{\rm F}$),
obtained by integrating the spectra in k-o within the
energy resolution window (10meV) about $E_{\rm F}$. (see Fig. 1b.). The three temperatures, $T_{\rm c}$ (green),
$T_{\rm pair}$ (red), and $T^*$ (blue) are indicated with dashed lines. The paring
weight (Wpair, red area) and the pseudogap weight ($W_{\rm PG}$, blue area) are
separated by a line extrapolated from the linear behavior of $W(E_{\rm F}$) at high
temperatures. }
\label{fig2}
\end{figure*}

\begin{figure*}[tbp]
\includegraphics[width=5in]{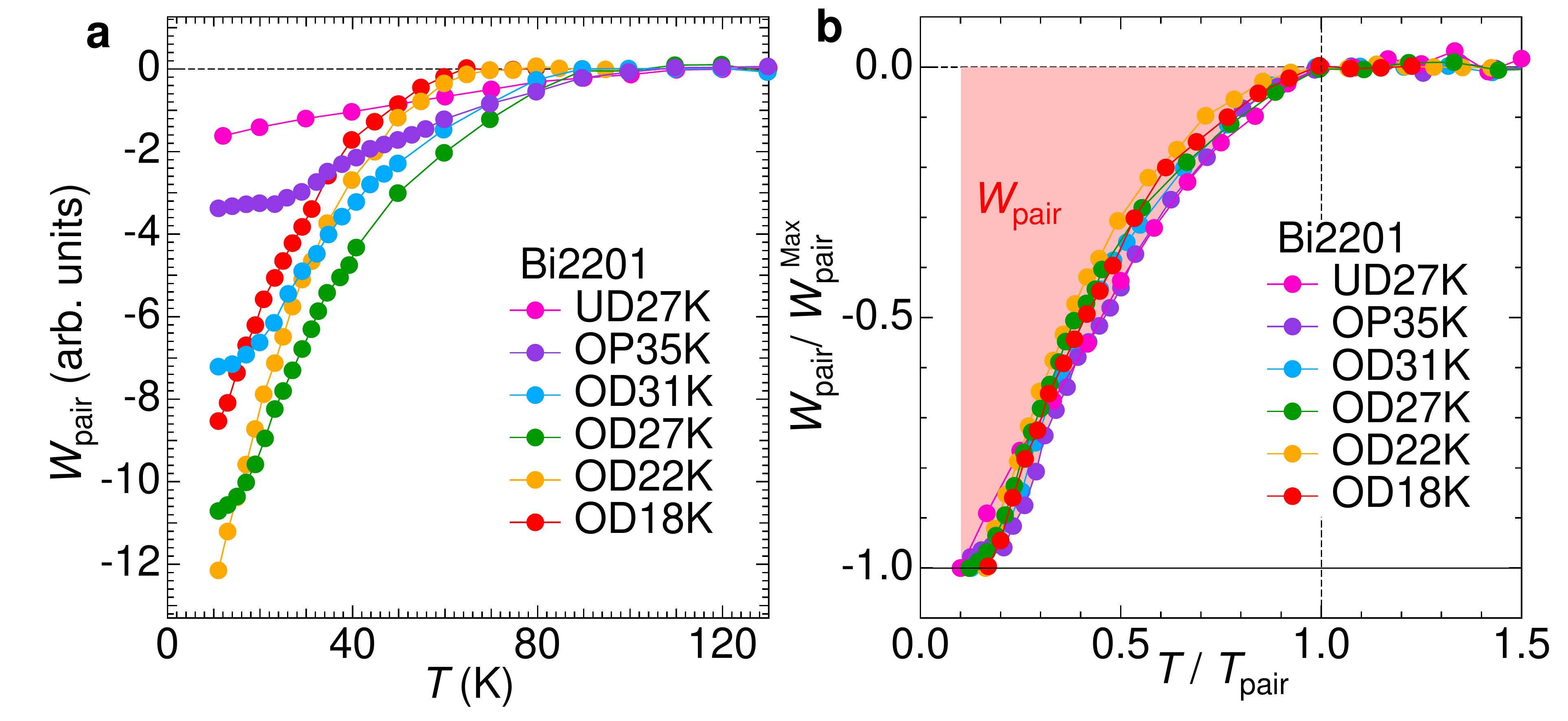}
\caption{Universal scaling behavior of the pairing spectral weight. {\bf a} Temperature dependence of ($W_{\rm pair}$) for
all samples extracted by subtracting extrapolated $W_PG$) line from $W(E_{\rm F}$) curves in Fig. 2(h-n). {\bf b} $W_{\rm pair}$ from
(a) scaled with the paring temperature ($T_{\rm pair}$ ) and the maximum value at the
lowest temperature ($W_{\rm pair}^{\rm Max} $). }
\label{fig3}
\end{figure*}

\begin{figure*}[tbp]
\includegraphics[width=6in]{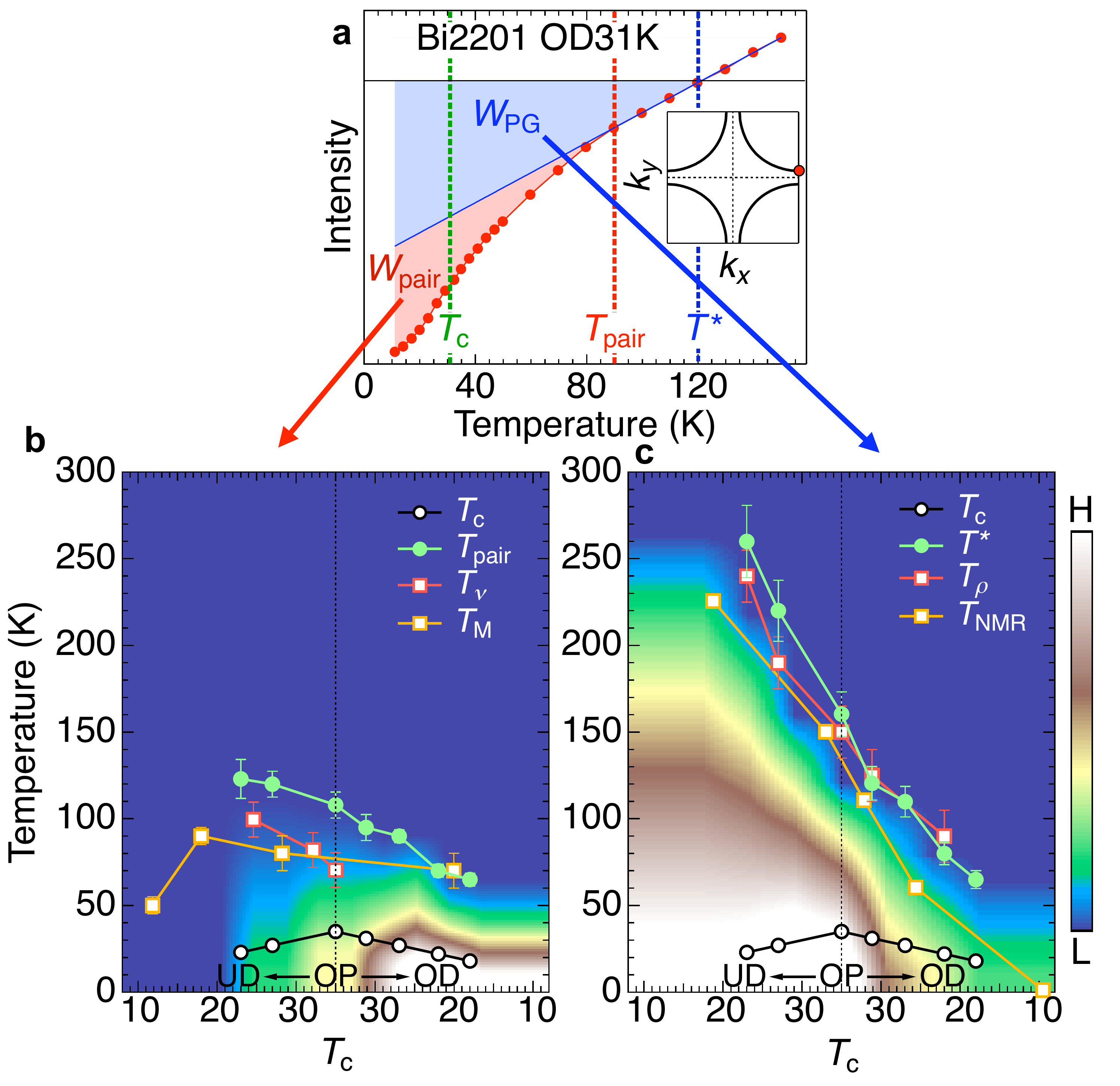}
\caption{Phase diagrams obtained using pairing and pseudogap spectral weights. {\bf a} $W(E_{\rm F}$) vs $T$ plots for OD31K, same as Fig. 2r. {\bf b} Phase
diagram obtained using the paring spectral weight ($W_{\rm pair}$, red
area in a). The onset temperature of pair formation estimated from our ARPES data ($T_{\rm pair}$),  the onset temperature of Nernst effect ($T_\upsilon$) \cite{OngPRB} and  diamagnetic effect ($T_{\rm M}$) \cite{OngDiamagnetism} are plotted. {\bf c} Phase diagram obtained using the pseudogap spectral weight ($W_{\rm PG}$, blue area in (a)). The pseudogap temperature estimated from our ARPES data ($T^*$), our resistivity results ($T_{\rm \rho }$, see Fig. S1) and NMR ($T_{\rm NMR}$) \cite{NMR} are plotted.}
\label{fig4}
\end{figure*}

\end{document}


\title{On-line supplementary information for

Disentangling Cooper-pair formation above $T_c$ from the pseudogap in cuprates}

\author{Takeshi~Kondo}
\affiliation{Ames Laboratory and Department of Physics and Astronomy, Iowa State
University, Ames, IA 50011, USA}

\author{Yoichiro~Hamaya}
\affiliation{Department of Crystalline Materials Science, Nagoya University, Nagoya
464-8603, Japan}

\author{Ari~Palczewski}
\affiliation{Ames Laboratory and Department of Physics and Astronomy, Iowa State
University, Ames, IA 50011, USA}

\author{Tsunehiro~Takeuchi}
\affiliation{Department of Crystalline Materials Science, Nagoya University, Nagoya
464-8603, Japan}
\affiliation{EcoTopia Science Institute, Nagoya University, Nagoya 464-8603, Japan}

\author{J.~S.~Wen}
\affiliation{Condensed Matter Physics and Materials Science Department, 
Brookhaven National Laboratory, Upton, New York 11973, USA}

\author{G.~Z.~J.~Xu}
\affiliation{Condensed Matter Physics and Materials Science Department, 
Brookhaven National Laboratory, Upton, New York 11973, USA}

\author{Genda~Gu}
\affiliation{Condensed Matter Physics and Materials Science Department, 
Brookhaven National Laboratory, Upton, New York 11973, USA}

\author{J\"org~Schmalian}
\affiliation{Ames Laboratory and Department of Physics and Astronomy, Iowa State
University, Ames, IA 50011, USA}

\author{Adam~Kaminski}
\affiliation{Ames Laboratory and Department of Physics and Astronomy, Iowa State
University, Ames, IA 50011, USA}
\date{\today }

\date{\today}

\maketitle

\section {Samples} 
Optimally doped Bi$_2$Sr$_2$CaCu$_2$O$_{8+\delta}$ (Bi2212) single crystals with $T_{\rm c}$=90K (OP90K) and 
(Bi,Pb)$_2$(Sr,La)$_2$CuO$_{6+\delta}$ (Bi2201) single crystals with various $T_{\rm c}$'s were grown by the conventional 
floating-zone (FZ) technique.  The carrier concentration of Bi2201 was controlled by the partial substitution of La for Sr and subsequent annealing. We obtained over-, optimally-, and underdoped samples with $T_{\rm c}$=18K, 22K, 27K, 35K, 27K, and 23K (OD18K, OD22K, OD27K, OD31K, OP35K, UD27K, and UD23K),  respectively. The  nominal composition and annealing conditions of these Bi2201 samples are summarized 
in Table I.  

One drawback of studying the pseudogap in high $T_{\rm c}$ cuprates such as Bi2212 is the large energy  scale of superconducting gap ($\sim$40 meV  at optimal doping), which is comparable to one of the pseudogap.  This similarity makes it difficult to conclude whether or not these are separate energy scales that correspond to two different states.  We chose to study Bi2201 and  discuss its properties in most of this work, because Bi2201 has a low $T_{\rm c}$ of $\sim$35K, but large $T^*$ similar to that of Bi2212 at optimal doping, and this make it possible to investigate separately characteristics of the superconducting and pseudogap states. Bi2201 has additional advantage for our study:  it is a single layered cuprate (i.e. it has a single  CuO$_2$ plane per unit cell), therefore its spectra are free from bilayer-induced band  splitting. 
We partially substituted Pb for Bi for all doping samples to suppress the modulation in the BiO plane, 
which usually contaminates the ARPES signal: the outgoing photoelectrons are diffracted, 
creating multiple images of the band and Fermi surface that are shifted in momentum. 
The bilayer-free and modulation-free samples enabled us to precisely analyze the ARPES  spectra. 

We measured the temperature dependence of the resistivity along the $a-b$ planes ($\rho _{ab}$) for OD22K, OD31K, OP35K, UD27K, and UD23K, and plot these in Fig. S1a. The signature of the psudogap can be clearly observed in these data: resistivity is linear at high temperatures, but a change of the slope occurs at $T^*$.
In the heavily overdoped sample OD22K, the resistivity has a $T^2$ component, which makes the detection of the pseudogap using above method difficult. 
However, the change of curvature due to the pseudogap can be  easily detected examining first derivative of resistivity as demonstrated in the inset of panel a. 
$T^*$ is defined as the temperature at which the slope of $d\rho _{ab}/dT$ changes. 
We plot $T^*$ determined from the resistivity along with $T_{\rm c}$ for each sample in panel c. 
$T^*$ obtained from the resistivity measurements agrees well with one determined from ARPES data using ``single spectral peak criterion" described in main text.  Figure S1b shows magnetic susceptibility of all Bi2201 samples used in the paper. 
All samples show a sharp superconducting transition ($\sim$5K). 

\section {Experimental method of ARPES}   

ARPES data was acquired using a laboratory-based system consisting of a Scienta SES2002 electron analyzer and GammaData Helium UV lamp. All data were acquired using the HeI line with a photon energy of 21.2 eV. The angular resolution was $0.13^\circ$ and $\sim 0.5^\circ$ along and perpendicular to the direction of the analyzer slits, respectively.  
The energy corresponding to the chemical potential was determined from Fermi edge of polycrystalline Au reference in electrical contact with the sample. The energy resolution was set at $\sim10$meV - confirmed by measuring the energy width between the 90$\%$ and 10$\%$ of the Fermi edge from the same Au reference. 
Custom designed refocusing optics enabled us to accumulate  high statistics spectra in a short time without being affected by possible sample surface aging due to absorption or loss of oxygen. Special care was taken in the purification of helium gas that supplies the UV source to remove even the smallest trace of contaminants that could contribute to surface contamination. Typically no changes in the spectral lineshape of samples were observed in consecutive measurements over several days. The samples were cooled using a closed-cycle refrigerator. Measurements were performed on several samples and we confirmed all yielded consistent results. 

\section {Normalization of ARPES spectra} 

In the analysis we used symmetrized EDCs normalized over the whole energy range (-0.4eV $\leq E\leq $ 0.4eV) for each spectrum.  Here we demonstrate that a particular normalization scheme does not affect the results of our analysis.
In Fig. S2c, we compare four $W(E_{\rm F}$)) vs $T$ curves from a Bi2212 OP90K sample obtained with four different schemes of the normalization. 
First one shows the result of the spectra normalized to the total area in the range of  -0.4eV to 0.4eV, as used in the the data analysis. 
The other three are normalized by the intensity within $\pm$30 meV centered at -0.4eV,  -0.3eV, and -0.2 eV.
All  $W(E_{\rm F}$) vs $T$ curves are very similar, with deviation from linear behavior occurring always at the same temperature of $\sim$150K. This allows us to conclude that presented results, including the determination of the pairing temperature do not dependent on particular normalization scheme. 
We have also verified in Fig. S2d that a energy width of integration ($\delta E$, see panel b) does not significantly affect the lineshape of  $W(E_{\rm F}$) vs $T$ curves. 
These checks demonstrate that our analysis extracted intrinsic behavior of the low energy electronic states which is independent of details of employed analysis.

\section { Spectral weight over a wide range of temperature}

In the  main text, we presented the data,  where the spectral weight, $W(E_{\rm F}$) displays a linear behavior even for temperatures slightly above $T^*$. Here we discuss the details of behavior above $T^*$. We also demonstrate that thanks to high quality data we can detect changes of the spectral gap across $T_{\rm c}$, which is strong evidence supporting existence its two components.
Figure S3a and b shows symmetrized EDCs plotted with and without offsets, respectively, measured over a wide range of temperatures from below $T_{\rm c}$ to far above $T^*$. The data were measured at the antinodal Fermi momentum for OP35K Bi2201 sample.  The panel c shows the same spectra as in panel b from which data at the highest temperature was subtracted. The spectral weight close to $E_{\rm F}$ ($W(E_{\rm F}$), hatched area) is plotted in panel d as a function of temperature. As discussed in the paper, the $W(E_{\rm F}$) increases linearly from $T_{\rm pair}$ to slightly above $T^*$. At higher temperatures, the curve deviates from the straight line, and  starts to decrease after reaching the maximum. We have defined $T^*$ as a temperature at which the two peaks in symmetrized EDC merge into one peak at elevated temperature, as it is customary in the literature and which leads to $T^*$ that agrees with one extracted from resistivity measurement.  The behavior of  $W(E_{\rm F}$) vs $T$ plot in panel d indicates that the spectral weight lost due to the pseudogap state ($W_{\rm PG}$) keeps filling up at $E_{\rm F}$ even slightly above the $T^*$, which is consistent with $T^*$ being a cross over temperature. At higher temperatures $W_{\rm PG}$ saturates and eventually it starts to decrease. This is an expected behavior for metallic samples, where thermal effects became dominant and broaden the spectral peak at $E_{\rm F}$, causing the peak height to decrease. In panel e, we plot the spectral gap (energy of the spectral peak indicated in panel a with bars) as a function of temperature. This very high quality data reveals for the first time a non-monotonic behavior; the spectral gap first increases across $T_{\rm c}$ , then decreases all the way down to zero at $T^*$. This characteristic behavior is consistent with the existence of two different spectral components: a smaller one due to pairing below $T_{\rm pair}$ and a larger one due to the pseudogap state.  The spectral gap is a convolution of the two energy scales and as a consequence it behaves abnormally with temperature.





\begin{table*}
\caption{Sample preparation conditions for the present Bi2201 single crystals.}
\begin{center}
   \begin{tabular}{c|c|c|c|c}
    \hline
    \hline    
     nominal composition & atmosphere & temperature & $T_{\rm c}$(K) & label  \\
    \hline     
     (Bi$_{1.74}$Pb$_{0.38}$)Sr$_{1.88}$CuO$_{6+\delta}$ & vacuum & 550C$^\circ $ for 24hours &  18  & OD18K \\
     (Bi$_{1.74}$Pb$_{0.38}$)Sr$_{1.88}$CuO$_{6+\delta}$ & vacuum & 650C$^\circ $ for 24hours &  22  & OD22K \\
     (Bi$_{1.35}$Pb$_{0.85}$)(Sr$_{1.47}$La$_{0.38}$)CuO$_{6+\delta}$ & Air & 600C$^\circ $ for 24hours & 27  & OD27K \\
     (Bi$_{1.35}$Pb$_{0.85}$)(Sr$_{1.47}$La$_{0.38}$)CuO$_{6+\delta}$ & Air & 450C$^\circ $ for 24hours & 31  & OD31K \\
     (Bi$_{1.35}$Pb$_{0.85}$)(Sr$_{1.47}$La$_{0.38}$)CuO$_{6+\delta}$ & Ar flow & 650C$^\circ $ for 24hours & 35 & OP35K \\
     (Bi$_{1.35}$Pb$_{0.85}$)(Sr$_{1.40}$La$_{0.45}$)CuO$_{6+\delta}$ & Ar flow & 650C$^\circ $ for 24hours & 27  & UD27K \\
     (Bi$_{1.1}$Pb$_{1.0}$)(Sr$_{1.28}$La$_{0.6}$)CuO$_{6+\delta}$ & vacuum & 650C$^\circ $ for 24hours & 23 & UD23K  \\
    \hline
    \hline
\end{tabular}
\end{center}
\end{table*}

\renewcommand{\thefigure}{S\arabic{figure}}
\begin{figure*}
\includegraphics[width=6.5in]{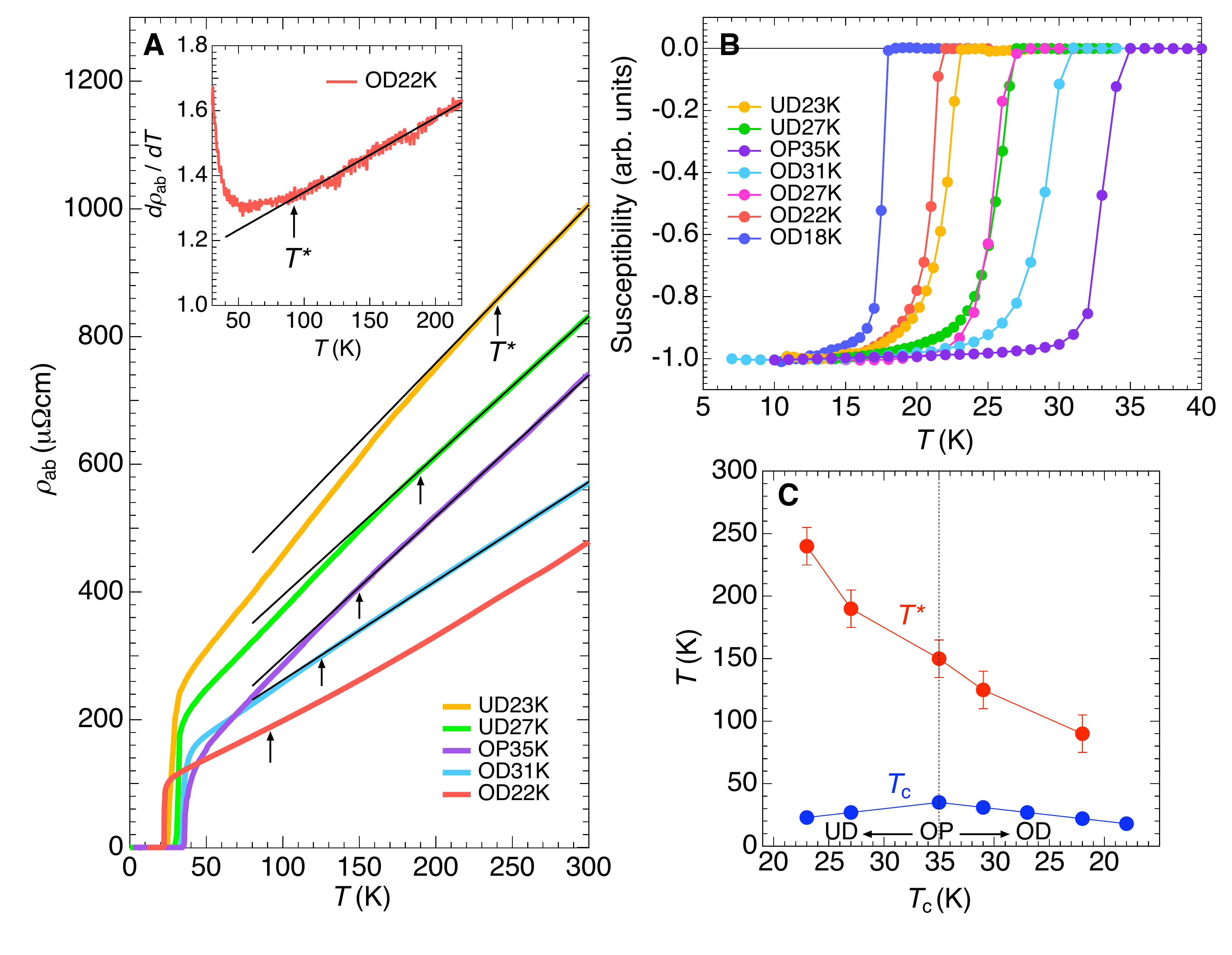}
\caption{Characterization of the samples using transport and susceptibility measurements.  {\bf a} Temperature dependence of resistivity along $a-b$ plane for UD23K, UD27K, OP35K, OD31K, and OD22K. Black lines are fitted to the resistivity at high temperatures. The arrow indicates the pseudogap temperature $T^*$ defined as a temperature at which the resistivity deviates from linear behavior at high temperatures. For OD22K, $T^*$ was estimated as a temperature at which  derivative of resistivity deviates from 
linear behavior. {\bf b} Magnetic susceptibility for all Bi2201 samples used in the paper. 
{\bf c} Phase diagram of Bi2201. $T_{\rm c}$ and $T^*$ are from analysis shown in panel a.}
\label{fig1}
\end{figure*}

\begin{figure*}
\includegraphics[width=6.5in]{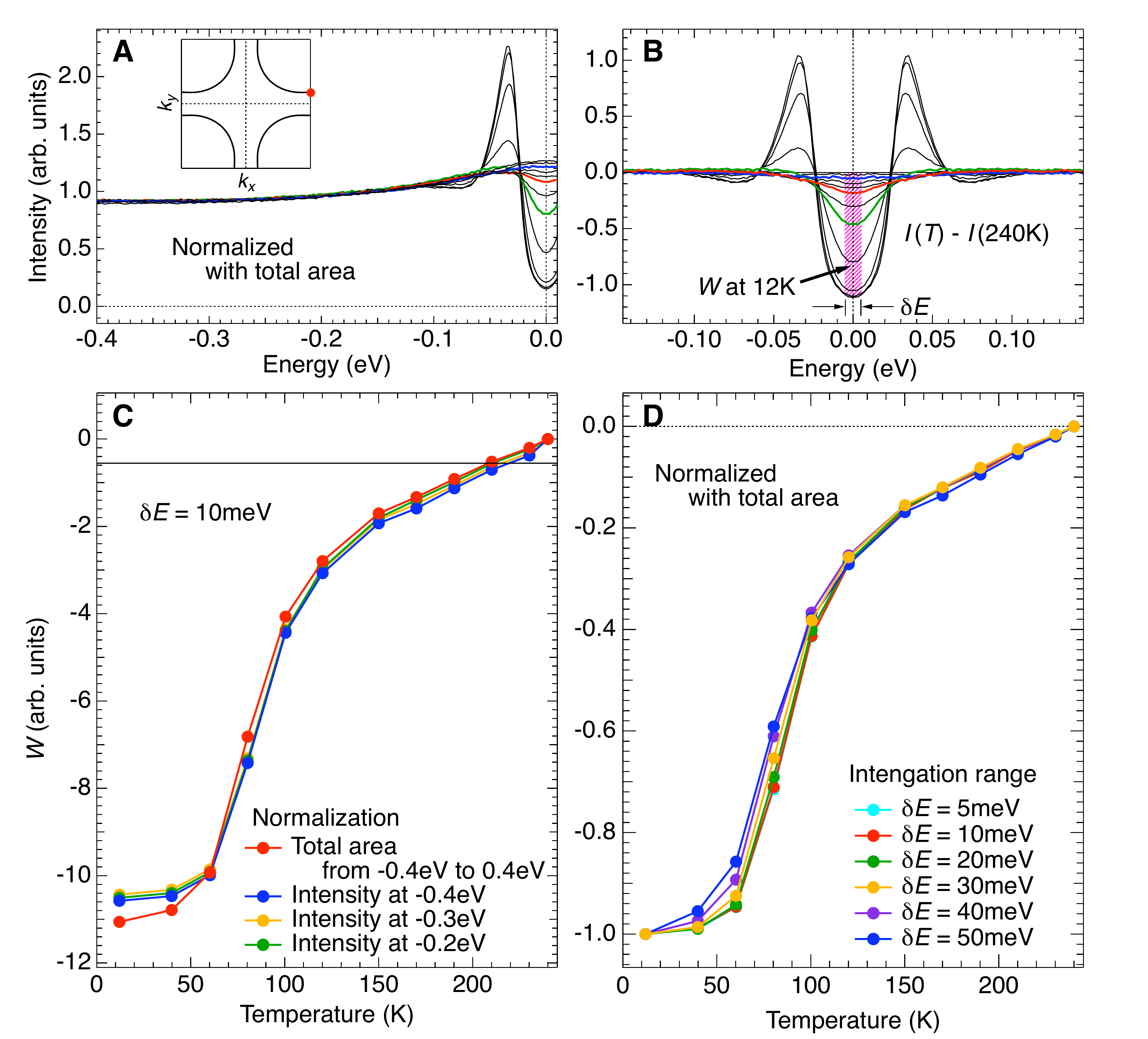}
\caption{Verification of normalization procedure.  {\bf a} Symmetrized EDCs at the antinodal Fermi momentum measured over a wide temperature range from below $T_{\rm c}$ to above $T^*$. All spectra are normalized to the total area integrated from -0.4eV to 0.4eV. {\bf b} Same spectra as in a subtracted by a spectrum at the highest temperature (240K). Spectral weight at $E_{\rm F}$, $W(E_{\rm F}$), is defined as an integrated area within an energy window of $dE$ centered at $E_{\rm F}$ (marked as pink, hatched box).  {\bf c} Temperature dependence of $W(E_{\rm F}$) for spectra normalized using four different normalization schemes: using total area of symmetrized EDCs in range of -0.4eV to 0.4eV and  normalized to the intensity within a narrow energy window ($\pm$30meV) centered at -0.4eV, -0.3eV, and -0.1eV.  {\bf d} Temperature dependence of $W(E_{\rm F}$) calculated using different energy windows: $dE$=5meV, 10meV, 20meV, 30meV, 40meV, and 50meV centered at $E_{\rm F}$.} 
\label{fig1}
\end{figure*}

\begin{figure*}
\includegraphics[width=3in]{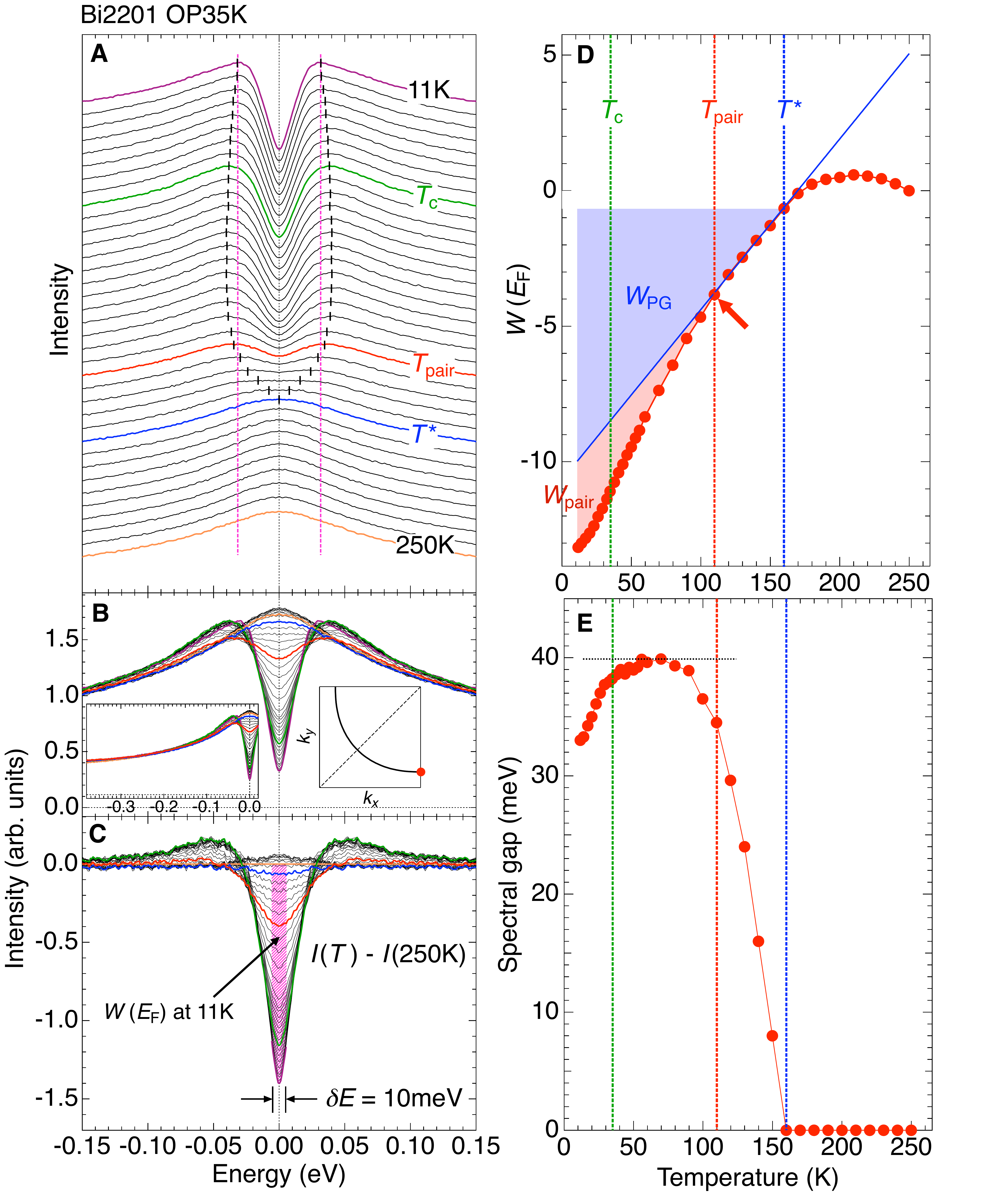}
\caption{{\bf a} Energy distribution curves at various temperatures from deep
below $T_{\rm c}$ to above the pseudogap closing temperature ($T^*$) measured at the
antinodal Fermi momentum in OP35K Bi2201. All spectra are
symmetrized about the Fermi level to remove the effect of the Fermi function, which cuts off the low energy intensity. Spectral gaps are indicated with bars. 
{\bf b} Same spectra as in A without offsets. 
{\bf c} Difference spectra: the spectrum at the highest temperature subtracted from each of spectra in b. 
These curves demonstrate the energy transfer of spectral weight from the high
temperature. The spectral weight close to the Fermi level ($W(E_{\rm F}$)) is
estimated by integrating the spectral intensity in c within an energy
window of the experimental energy resolution (10meV). 
{\bf d} The temperature dependence of $W(E_{\rm F}$), estimated from spectra in c. The paring temperature,
$T_{\rm pair} $ , is defined as the onset temperature of deviation in $W(E_{\rm F}$) vs $T$ curve
from a linear behavior at higher temperatures. The pseudogap closing
temperature, $T^*$, is defined as the temperature where the two spectral peaks in
the symmetrized EDCs merge into a single peak as seen in a. Three temperatures, $T_{\rm c}$,
$T_{\rm pair} $ , and $T^*$ are indicated with dashed lines. Two components of the paring
weight ($W_{\rm pair}$, red area) and the pseudogap weight ($W_{\rm PG}$, blue area) are
separated by a line extrapolated from the linear behavior of $W(E_{\rm F}$) at high
temperatures. 
{\bf e} Temperature dependence of spectral gap estimated from the peak energy in symmetrized EDC (see black bars in ). The maximum of the energy gap is indicated by a dotted line.}
\label{fig1}
\end{figure*}